\documentclass[final,5p,times,twocolumn,sort&compress]{elsarticle}

\usepackage{amsmath,amssymb,graphicx}
\usepackage{enumitem}
\usepackage{amsbsy}
\usepackage{latexsym}
\usepackage{color}
\usepackage{graphicx}
\usepackage{psfrag}
\usepackage[normalem]{ulem}

\newcommand{\be}{\begin{equation}}
\newcommand{\ee}{\end{equation}}
\newcommand{\bea}{\begin{eqnarray}}
\newcommand{\eea}{\end{eqnarray}}

\newcommand{\comment}[1]{}

\begin{document}

\title{Domain statistics in the relaxation of the one-dimensional Ising model with strong long-range interactions}

\author{Federico Corberi}
\ead{corberi@sa.infn.it}
\address{Dipartimento di Fisica ``E. R. Caianiello'' and INFN, Gruppo Collegato di Salerno, via Giovanni Paolo II 132, 84084 Fisciano (SA), Italy.}
\author{Manoj Kumar \corref{cor1}}
\ead{manojkmr8788@gmail.com}
\cortext[cor1]{Corresponding author}

\address{Institut für Physik, Technische Universität Chemnitz, 09107 Chemnitz, Germany}

\address{Physics Department, GITAM School of Science, GITAM (Deemed to be University), Hyderabad, Telangana 502329, India.** }
\cortext[auth]{Present affiliation} 
\author{Eugenio Lippiello}
\ead{eugenio.lippiello@unicampania.it}
\address{Dipartimento di Matematica e Fisica, Universit\`a della Campania ``L. Vanvitelli'',
Viale Lincoln 5, 81100, Caserta, Italy}
\author{Paolo Politi}
\ead{paolo.politi@cnr.it}
\address{Istituto dei Sistemi Complessi, Consiglio Nazionale delle Ricerche, Via Madonna del Piano 10, 50019 Sesto Fiorentino, Italy}
\address{ INFN Sezione di Firenze, via G. Sansone 1, 50019 Sesto Fiorentino, Italy}

\begin{abstract}

After a zero temperature quench, we study the kinetics of the one-dimensional Ising model with long-range interactions between spins at distance $r$ decaying as $r^{-\alpha}$, with $\alpha \le 1$. 
As shown in our recent study [SciPost Phys 10, 109 (2021)] that only a fraction of the non-equilibrium trajectories is characterized by the presence of coarsening domains while in the remaining ones
the system is quickly driven towards a magnetised state. Restricting to realisations displaying coarsening we compute numerically the probability distribution of the size of
the domains and find that it exhibits a scaling behaviour with an unusual $\alpha$-dependent power-law decay. This peculiar behaviour is also related to the divergence of the average size of domains with system size at \textit{finite} times. Such a scenario differs from the one observed when $\alpha>1$,
where the distribution decays exponentially. 
Finally, based on numerical results and on analytical calculations we argue that the average domain size grows asymptotically linearly in time.

\end{abstract}

\maketitle

\section{Introduction}
\label{intro}

The phase-ordering kinetics of ferromagnetic systems is  perhaps the simplest non-trivial paradigm of 
slow non-equilibrium evolution~\cite{Bray94,Desai_Kapral,libro}.
After a deep temperature quench the system orders in a rather different way depending on the
range of the interactions.
When they decay sufficiently fast with the distance among constituents, the evolution proceeds by the formation and growth of domains of the low-temperature, symmetry-related, equilibrium phases. If the quench is done from above to below the critical temperature $T_c$, symmetry is never broken in the thermodynamic limit. Instead, in a finite system, it occurs only when the largest domains have grown comparable to the system's size, which we will denote in the following as a border finite-size effect. In the case of interactions of the form $J(r)\propto r^{-\alpha}$ ~\cite{CLP_review,Defenu_2020,iannone2021}, which will be considered in this paper, the phenomenology described above is observed in the so-called weak long-range case (WLR), namely when $\alpha >d$, where $d$ is the spatial physical dimension. This guarantees that some foundations
of usual statistical mechanics, such as additivity and extensivity, are preserved~\cite{book_long_range}.

In the opposite limit of mean-field systems, one observes a completely different route to order. In this case a feedback effect makes the tiny magnetisation of the initial state grow up fast to macroscopic values, rapidly determining the formation of a homogeneous phase without development of domains. Phase ordering terminates in a time of order one.

\begin{figure}[h]
	\centering
 \includegraphics[width=0.9\linewidth]{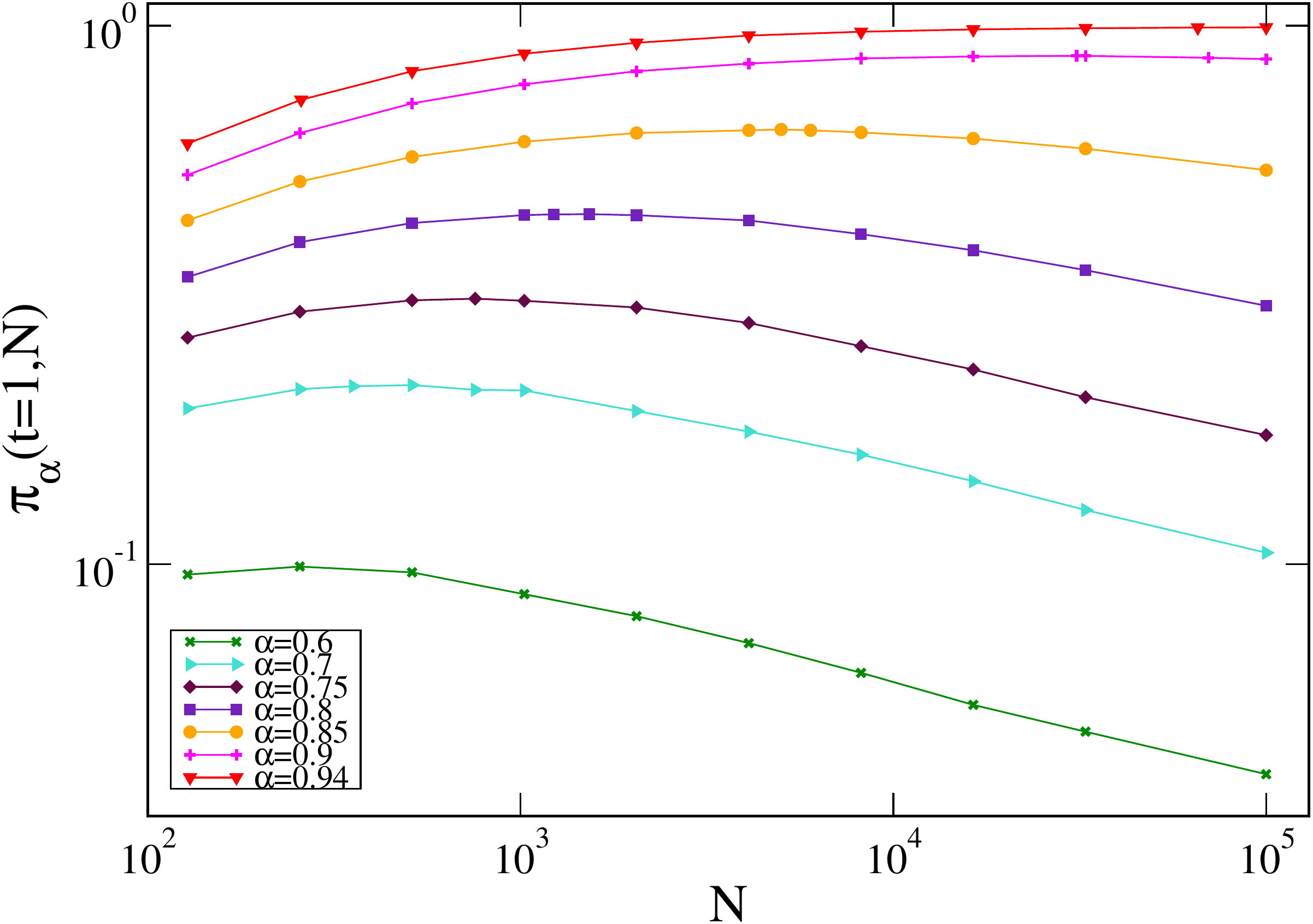}
 \caption{The fraction $\pi_\alpha(t=1,N)$ of realisations that, at time $t=1$ after a quench to zero temperature, display domains. Time $t=1$ is the   characteristic time where the mean-field
 evolution reaches the ordered state.
 }
	\label{fig_oldP}
\end{figure}

The evolution of systems with strong long-range interactions (SLR), i.e. with an algebraic decay of $J(r)$ and $\alpha \le d$, was recently studied in one dimension in~\cite{iannone2021}. 
There it was shown that, for a given system size, the non-equilibrium ensemble contains a fraction of trajectories along which coarsening domains are unexpectedly displayed, whereas the
remaining ones behave similarly to the mean field, as shown in Fig.~\ref{fig_oldP} for systems of different sizes $N$ and for different values of $\alpha$.

The relative abundance of these two kinds of trajectories depends on $N$,  and on $\alpha$.
Specifically, the number of the mean-field ones becomes more significative as $\alpha $ decreases (keeping $N$ fixed) or increasing $N$ for a given value of $\alpha$. A consequence of this radically different character of the two kinds of trajectories is the fact that averaging them altogether mixes different features and hinders the physical mechanisms at work.
Instead, in order to grasp the physics beyond the two kinds of processes one should look separately at the coarsening and at the mean-field-like configurations. 

This is the program we pursue in this paper. Specifically, after a zero temperature quench from the completely disordered state, we isolate the coarsening trajectories and study the statistics of the domains contained therein, computing numerically the probability distribution $p_\alpha (s,t)$ of observing domains of size $s$ after a time $t$ has elapsed since the quenching, for different values of $\alpha$. This quantity, for $\alpha \le 1$, exhibits an algebraic decay different from the exponential one observed for $\alpha>1$.  We find that $p_\alpha$ takes a scaling form, similar to what observed in WLR systems. This suggests that also in the SLR case the dynamical scaling symmetry informs the coarsening kinetics. However, beyond such similarity, many differences emerge, most prominently the major role played by finite-size effects, as compared to systems with WLR, whereby the characteristic domains' size depends on $N$ even in the thermodynamic limit $N\to \infty$. Combining the results 
of numerical simulations with analytical calculations, we argue that the typical domains size grows 
linearly in time, similar to what is found in systems of WLR quenched at zero temperature. 

In conclusion, the study undertaken in this paper shows that, if trajectories with domains are singled out, their kinetics is characterized by the same ingredients which are well known for zero-temperature quenches with WLR, namely dynamical scaling and ballistic growth~\footnote{We remind that for WLR the coarsening is ballistic for quenches to zero temperature while there is a slower algebraic growth~\cite{CLP_review} in quenches to finite temperatures.}.

This paper is organised into five sections. In the next one, we define the model and the quantities that will be considered further on. Sec.~\ref{results} presents and discusses the results of extensive
numerical simulations of the model. In particular we consider the scaling properties of the domain's probability distribution and the growth-law of the average domain's size. In Sec.~\ref{analytical} we present an analytical solution of the model, which is expected to be valid for long times in the thermodynamic limit. The growth of domains is predicted in this way to be linear in time.
Finally, in Sec.~\ref{concl} we recapitulate and point out some of the many open questions.

\section{The model} \label{model}
We consider a one-dimensional Ising model with $N$ spins $s_i=\pm 1$ on the sites of 
a linear lattice, described by the Hamiltonian
\begin{equation}
{\cal H}= -\sum _i s_i h_i,
\label{ham}
\end{equation}
where
\be
h_i\equiv \sum_{j\neq i}J_{ij}s_j
\label{locfield}
\ee
is the local field, or Weiss field. 
The coupling constant is
\be
J_{ij}=\frac{1}{2K(N)}\, \frac{1}{r_{ij}^{\alpha}},
\label{eqJ}
\ee
where $r_{ij}$ is the distance between two spins on the sites $i,j$ of the lattice and $K(N)$ is the Kac factor, namely
\begin{equation}
    K(N)=\sum_{r=1}^{N/2}\frac{1}{r^{\alpha}}.
\label{eqKac}
\end{equation}
This quantity is introduced to make the system extensive when $\alpha \le 1$, namely in
the SLR case we will focus in this paper. However this does not reinstate additivity since, for instance, the act of breaking a sample into two parts and bringing them far away changes the energy of the system also in the thermodynamic limit, as one can easily get convinced.
The distance $r_{ij}$ in Eq.~(\ref{eqJ}) is evaluated to take into account the periodic boundary conditions, namely
$r_{ij}=\min\{|i-j|, N-|i-j|\}$. 
The properties of the usual nearest neighbour model  $J_{ij}=\delta_{i\pm 1,j}$ are recovered letting $\alpha \to \infty$.

The equilibrium states of this system do not exhibit long-range order at finite temperatures as long as $\alpha >2$ ~\cite{Peierls1934,Dyson1969,Frohlich1982,Imbrie1988,Luijten2001,Mukamel2009}.
For $\alpha <2$, instead, there is a second-order
phase transition at finite critical temperature $T_c$. 
Right at $\alpha =2$ a Kosterlitz-Thouless phase transition occurs characterised by a jump of the magnetisation. $\alpha =0$ is the case of mean field, whose critical exponents remain unchanged up
to $\alpha =3/2$~\cite{Mukamel2009}. 

In a magnetic system, where the evolution does not conserve the order parameter, single spins $s_i$ are
randomly chosen and reversed with a transition rate $w(s_i)$ obeying detailed balance,
namely $w(s_i)/w(-s_i)=e^{-\beta (H_a-H_b)}$, where $H_b$ and $H_a$ are the energies of the system 
before and after the elementary move and $\beta $ is the inverse temperature, $\beta=1/(k_B T)$, where $k_B$ is the Boltzmann constant that we will set to unity in the following. We will consider Glauber transition rates
\be
w(s_i)=\frac{1}{2}\left [1-s_i\tanh(\beta h_i)\right ].
\label{GlauberRates}
\ee

The quenching procedure we consider amounts to preparing the magnet  in an equilibrium state at the initial temperature $T_i=\infty$, where spins are uncorrelated, and then instantly cooling it down to $T=0$ at time $t=0$. Notice that the evolution at zero temperature proceeds by randomly choosing single spins and flipping them if they are antiparallel to the local field. This will be important in the following.

In a previous article~\cite{iannone2021} we have shown that for $\alpha \le 1$, the presence of configurations with coarsening domains is a stochastic phenomenon which may occur (or not), depending on the different dynamical realisations, with a given probability $\pi_\alpha(t,N)$ plotted in Fig.~\ref{fig_oldP} and defined as the fraction of trajectories that at time $t=1$ presents at least two domains. Time is measured in units of MCS. In this paper we will restrict our attention only to those configurations where, at the observation time $t$, domains are present. In doing that, the ensemble of
trajectories we keep under study changes in time. Let us clarify that, from now on, the world {\it domain} does not refer to spin domains, but to local field domains, i.e. regions of the lattice with consecutive local fields $h_i$ of the same sign. Using local field domains is more physical because it neglects fast fluctuations of individual spins. Indeed a small fluctuation causing the flip of a spin inside a spin domain breaks it into two pieces, whereas field domains are not affected by this kind of fluctuation. Let us notice that at the quench temperature $T=0$ considered in this
paper, field and spin domains almost coincide because spins align in a time of order one with their local field, hence field domains and spin domains provide basically the same information. The situation may change at finite temperatures.

Defining the field-interfaces as sites $i$ where $h_ih_{i+1}<0$ and $s$, the size of the domain, as the distance between two consecutive such interfaces, we consider the probability $p_\alpha(s,t,N)$ of finding a domain of size $s$ at time $t$, in a system of total size $N$~\footnote{Notice that on the configurations with domains we are focusing on, it is $p_\alpha(s=N,t,N)\equiv0$, by definition.}. Numerically this is computed by quenching the model, following its evolution
up to time $t$ and then, if domains are present in the sample, counting the fraction of them with a given
size $s$. This quantity is then averaged over many different initial conditions sampled from the equilibrium state at infinite temperature, and over different thermal histories, namely over the various stochastic realisations of the kinetic evolution.

From $p_\alpha (s,t,N)$ the average domains size $\langle s_\alpha(t,N)\rangle$ can be extracted as
\be
\langle s_\alpha(t,N)\rangle = \sum _{s=1}^Ns\,p_\alpha(s,t,N).
\ee
Notice that this quantity may be ill-defined in the thermodynamic limit if $p_\alpha$ does not decay 
sufficiently fast at large $s$, a fact that actually occurs, as we will see, for $\alpha\le 1$.
However, for any finite $N$, the definition of $\langle s_\alpha(t,N)\rangle$ is sound.

In this paper we will consider only values $\alpha\ge0.8$. For smaller values of $\alpha$ it is difficult to have clear-cut evidence due to severe finite-size effects, as will be discussed further on.

\section{Results} \label{results}

\subsection{Overall behaviour of $p_\alpha$} \label{overall}

\begin{figure}[h]
\centering
	\rotatebox{0}{\resizebox{0.45\textwidth}{!}{\includegraphics{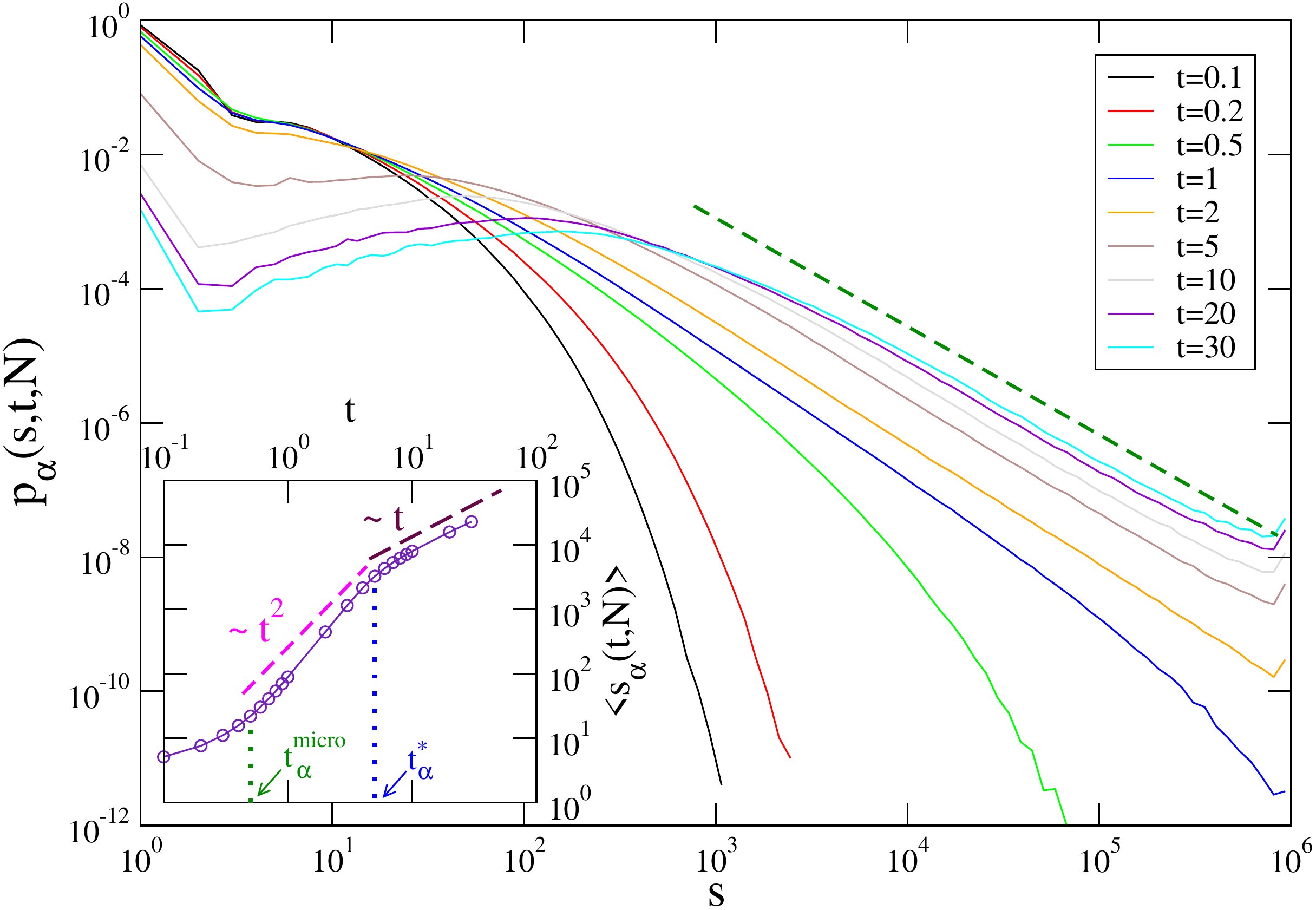}}}\vspace{0.3cm}
 \rotatebox{0}{\resizebox{0.45\textwidth}{!}{\includegraphics{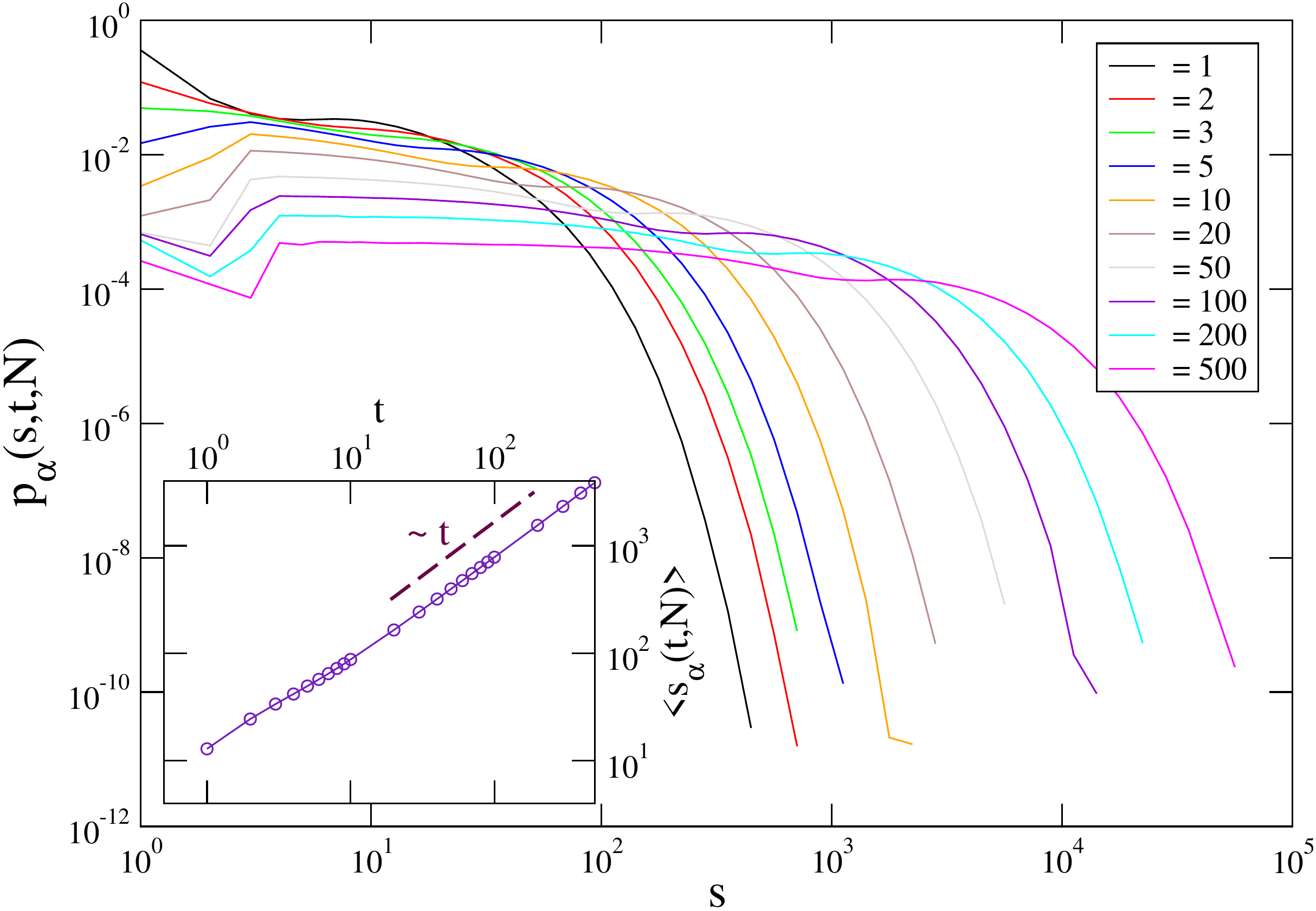}}}
 \caption{$p_\alpha(s,t,N)$ is plotted against $s$, for $\alpha=0.95$ (upper panel) and $\alpha=1.5$ (lower panel), $N=10^6$, and different times (see key). The green bold-dashed line in upper panel
 is the behavior $\propto s^{-(2-b_\alpha)}$, with $b_\alpha=0.387$. In the inset the average domains size $\langle s_\alpha (t,N)\rangle$ is plotted against time.
 The magenta and maroon bold-dashed lines are the behaviors $\propto t^2$ and $\propto t$, respectively.}
	\label{fig_p095}
\end{figure}

Before discussing the general features of $p_\alpha(s,t,N)$ we present here a comparison between a case with SLR ($\alpha=0.95$) and one case with WLR ($\alpha =1.5$). These two cases are portrayed 
in Fig.~\ref{fig_p095}. As we will see, the behavior of SLR system is qualitatively similar to the one presented here for all the values of $\alpha$ considered.

For WLR (lower panel), $p_\alpha$ has an exponential-like tail at large $s$ and the effect of increasing time is to move to larger values of $s$ the typical cut-off size. Consequently, $\langle s_\alpha(t,N)\rangle$ increases in an approximately linear way (inset in lower panel), as one expects for any $\alpha >1$~\cite{CLP_review,CLP_epl,CLP_lambda}, at $T=0$.

For $\alpha \le 1$, instead, the evolution can be divided into different time regimes. At sufficiently short times $t<t^{\rm micro}_\alpha$ ($t^{\rm micro}_\alpha\sim 0.5$ for the case shown in Fig.~\ref{fig_p095}) $p_\alpha$ has an exponential tail and the cut-off value increases with time, similarly to what happens for $\alpha >1$. At variance with the WLR case, however, by increasing $t$, $p_\alpha$ gradually builds-up an algebraic decay $s^{-(2-b_\alpha)}$, with $b_\alpha \simeq 0.39$, covering an increasing range of $s$-values starting at $s\simeq 10$ up to the point where the exponential cutoff sets in. Notice that such value of the exponent $b$ implies that all the moments of the distribution cannot be defined in the thermodynamic limit, including $\langle s_\alpha(t,N)\rangle$. We will see that this is true for any $\alpha \le 1$. Right at $t=t^{\rm micro}_\alpha$ the exponential cutoff reaches the largest available size and the power-law decay of $p_\alpha$ is approximately obeyed for any $s\gtrsim 10$. From this time on, we expect strong finite size effects to characterize the distribution in the region $s\lesssim N$.
Configurations with domains of such sizes are expected to become fully ordered in a very short time and to exit from the statistics of the domains configuration we are considering here. Finite size effects will be further discussed in Sec.~\ref{finites}. After $t^{\rm micro}_\alpha$ there are not important qualitative differences up to a later characteristic time $t^*_\alpha$ ($t^*_\alpha\simeq 5$ for the case shown in Fig.~\ref{fig_p095}), the only observed effect being the appearance of a spike at $s\simeq N$, signalling that domains almost as large as the whole system have been formed. In the time domain with $t>t^*_\alpha$, $p_\alpha$ develops a dip around $s\simeq 2$. What happens here is that the presence of very large domains suppresses the smallest ones. The effect of this is to generate a maximum at intermediate values of $s$, which moves towards larger $s$-values as time elapses. In the following we will call the two regimes discussed above, with $t^{\rm micro}_\alpha<t<t^*_\alpha$ and with $t>t^*_\alpha$ as regime I and II, respectively. 

The growth of $\langle s_\alpha (t,N)\rangle$ itself reflects the presence of these two different regimes, as it can be appreciated in the inset of the upper panel of Fig.~\ref{fig_p095}. Indeed, in regime I, $\langle s_\alpha(t,N)\rangle$ increases as $t^{\frac{1}{z}}$, with $z\simeq \frac{1}{2}$. Instead, in regime II, $\langle s_\alpha(t,N)\rangle$ keeps growing algebraically in time, but with a different exponent that, for the case of Fig.~\ref{fig_p095}, is of order $z=1$. We recall that 
$z=1$ is observed, for zero-temperature quenches, also with WLR.

Let us mention that, as we will discuss in the rest of the paper, $t^{\rm micro}_\alpha$ and $t^*_\alpha$ appear to be approximately $N$-independent. Therefore regime I cannot be expanded beyond the few 
time units separating $t^{\rm micro}_\alpha$ from $t^*_\alpha$. Instead, as we will clarify better later on,
regime II widens in time upon increasing $N$.

\subsection{Scaling} \label{scaling}

\begin{figure}[h]
\centering
	\rotatebox{0}{\resizebox{0.45\textwidth}{!}{\includegraphics{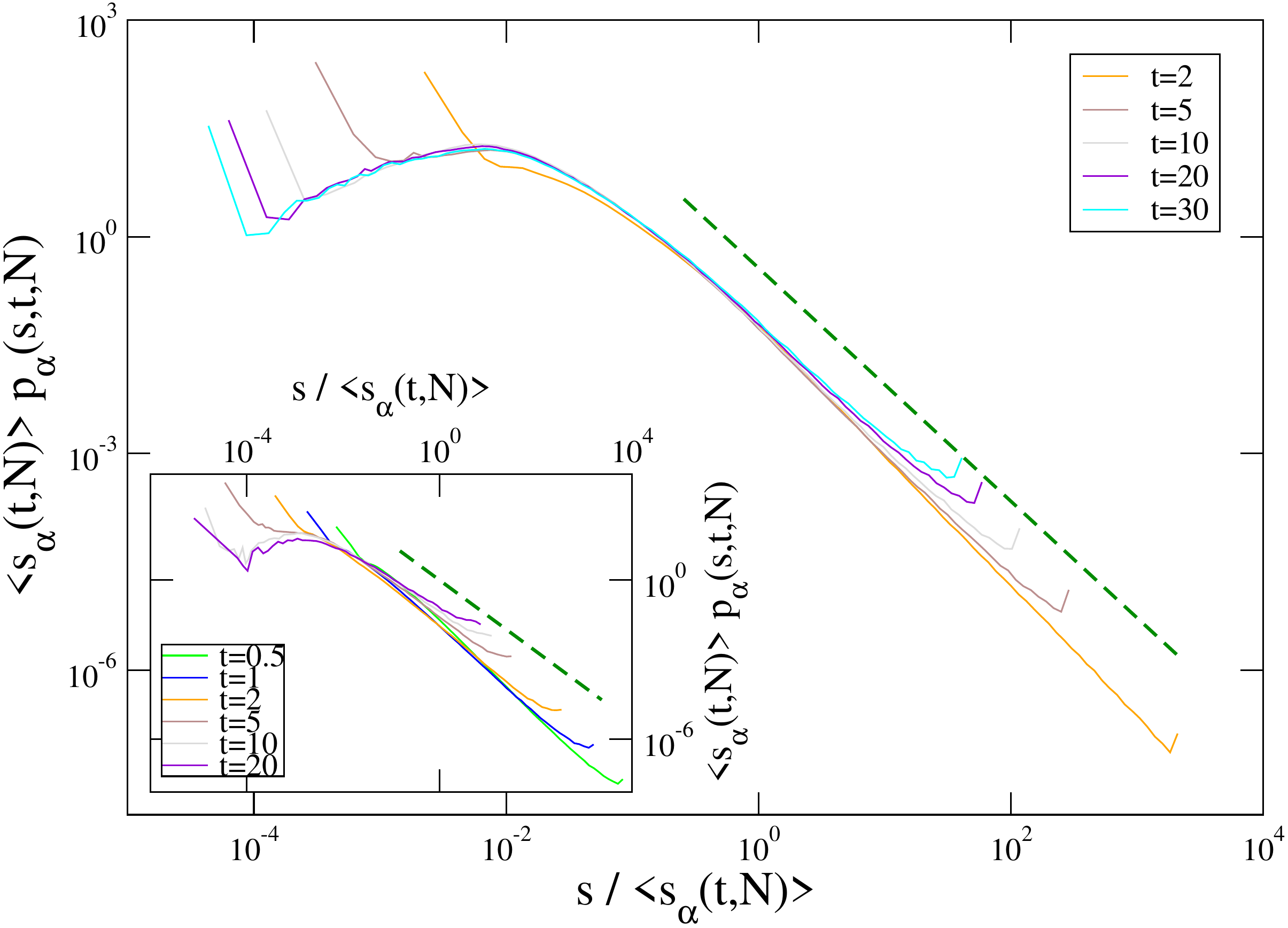}}}\vspace{0.3cm}
 \rotatebox{0}{\resizebox{0.45\textwidth}{!}{\includegraphics{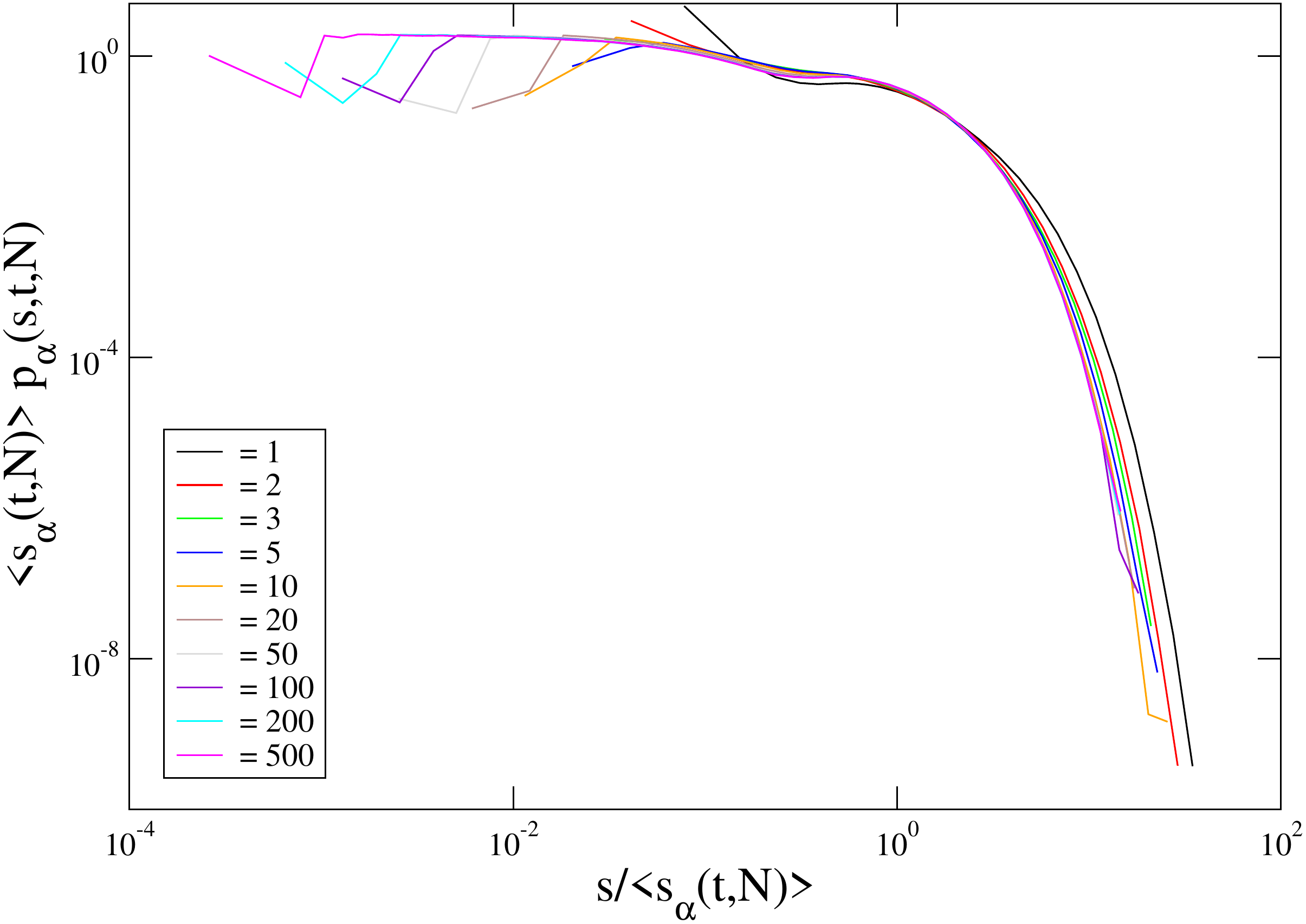}}}
 \caption{$\langle s_\alpha(t,N)\rangle \,p_\alpha(s,t,N)$ is plotted against $\frac{s}{\langle s_\alpha(t,N)\rangle}$, for $\alpha=0.95$ (upper panel) and $\alpha=1.5$ (lower panel), $N=10^6$, and different times (see key). The green bold-dashed line in upper panel is the behaviour $\propto s^{-(2-b_\alpha)}$, with $b_\alpha=0.387$. The inset shows the case with $\alpha=0.85$. In this
 case the green bold-dashed line is drawn with $b_\alpha=0.671$.}
	\label{fig_p095_scaled}
\end{figure}

It is well known~\cite{ALEMANY199518,PhysRevE.54.2513,Ben-Naim1998} that in the WLR case dynamical scaling holds, meaning that
\be
p_\alpha(s,t,N)=\langle s_\alpha(t,N)\rangle^{-1} \,g\left (\frac {s}{\langle s_\alpha (t,N)\rangle}\right ),\quad \mbox{for fixed }N,
\label{scalpt}
\ee
where $g$ is a scaling function. This expresses the fact that a single length $\langle s_\alpha(t,N)\rangle$ exists in the system
such that configurations at different times are statistically equivalent if lengths, as $s$, are measured in units of 
$\langle s_\alpha (t,N)\rangle$. The extra factor in front of $g$ is needed to preserve the normalisation of probability.
The occurrence of such scaling form is observed for $\alpha =1.5$ in the lower panel of Fig.~\ref{fig_p095_scaled}, 
with only small preasymptotic corrections at early times.
A similar plot is shown for SLR in the upper panel, for $\alpha =0.95$, restricting to times $t>t^{\rm micro}_{\alpha}$.
This shows that a scaling symmetry as in Eq.~(\ref{scalpt}) is obeyed also in the configurations containing domains 
for SLR. Notice that, for the time being, we 
test scaling with a fixed value of $N$ while we will consider the effect of changing $N$ in 
Sec.~\ref{finites}.
An analogous scaling is observed for any considered value of $\alpha$, see for instance the inset of the upper panel of Fig.~\ref{fig_p095_scaled}
for the case $\alpha=0.85$. Notice that, as time goes on, scaling is progressively extending to lower values of $\frac{s}{\langle s_\alpha(t,N)\rangle}$ as long as the dip at small $s$, discussed above, is formed. On the other hand, for large 
$\frac{s}{\langle s_\alpha(t,N)\rangle}$ scaling is correspondingly suppressed, due to the formation of very large domains. The largest domains, indeed, do not obey the scaling symmetry~(\ref{scalpt}), as it is expected because, as already anticipated, they are so much affected by finite-size effects.

\subsection{Finite size effects} \label{finites}

\begin{figure}[h]
	\centering
	\rotatebox{0}{\resizebox{0.45\textwidth}{!}{\includegraphics{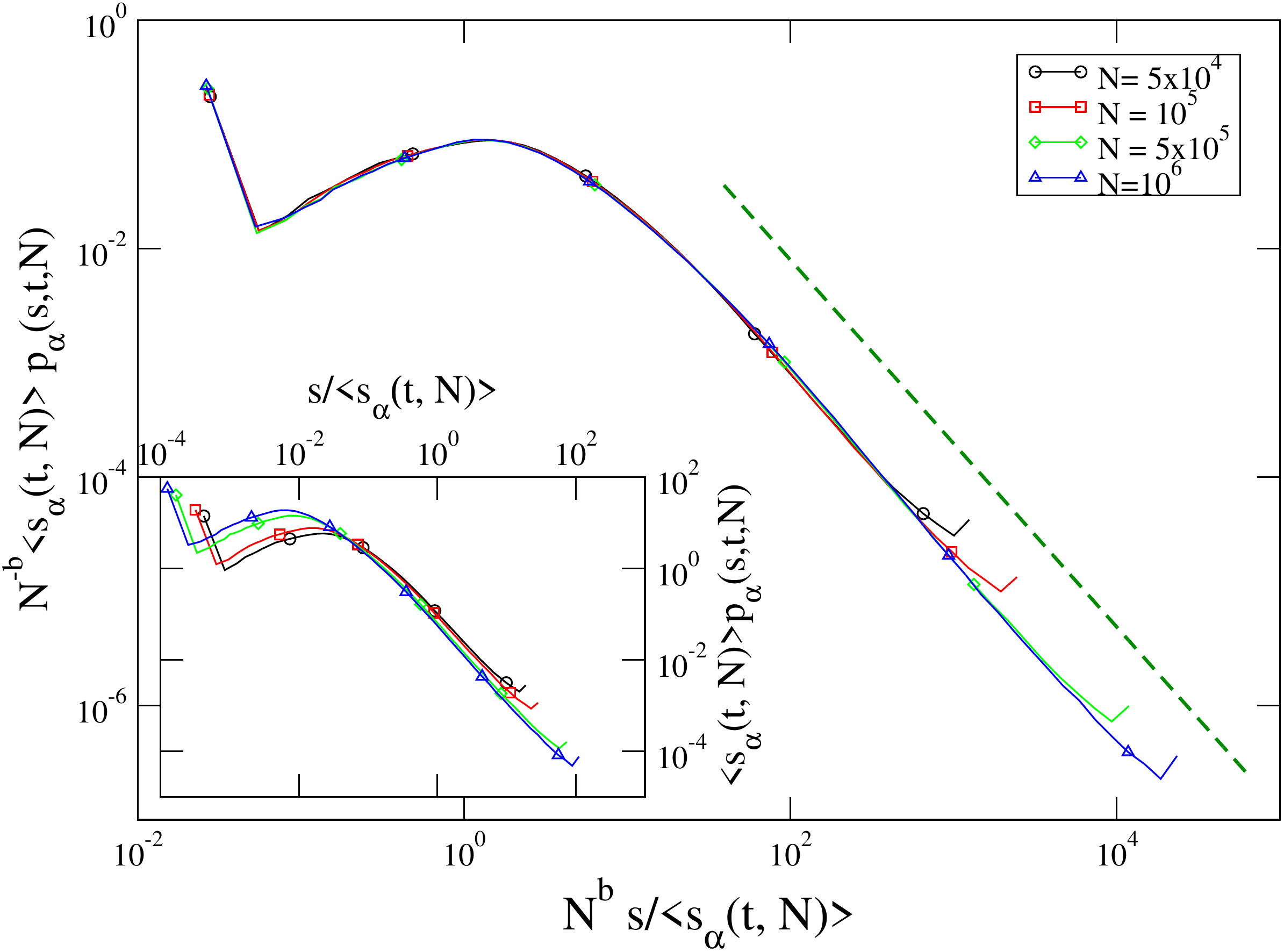}}}\vspace{0.3cm}
 \rotatebox{0}{\resizebox{0.45\textwidth}{!}{\includegraphics{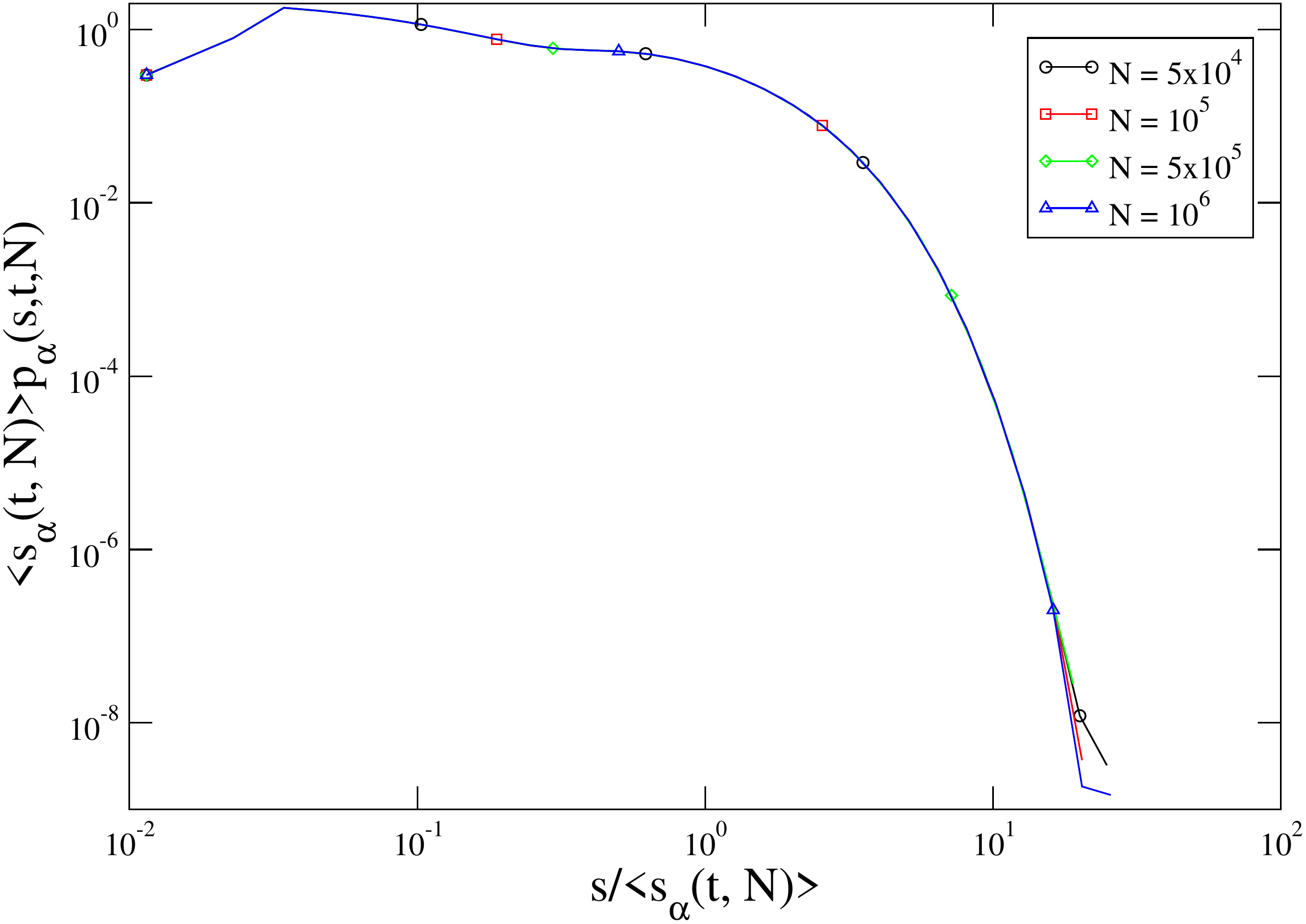}}}
 \caption{In the upper panel $N^{-b_\alpha} \langle s_\alpha(t,N)\rangle \,p_\alpha(s,t,N)$ is plotted against $N^{b_\alpha} \frac{s}{\langle s_\alpha(t,N)\rangle}$, for $t=10$, $\alpha=0.95$ and different system sizes $N$ (see key). The green dashed line is the behaviour $\propto x^{-(2-b_\alpha)}$, with $b_\alpha=0.387$. In the inset, we show the same set of data without any $N$-rescaling, namely plotting $p_\alpha(s,t,N)$ against $\frac{s}{\langle s_\alpha(t,N)\rangle}$.
 On the lower panel $p_\alpha(s,t,N)$
 is plotted against $s$, for $t=10$,
 $\alpha =1.5$ and different system sizes $N$ (see key).}
	\label{fig_p095_scaled_withN}
\end{figure}

In the WLR case, finite size effects become manifest when $\langle s_\alpha(t,N)\rangle$ becomes of the order of the system size $N$.
We will call this a {\it border finite-size effect} because it occurs when domains feel the system's border (even if, with the periodic boundary conditions we are using, there is not a real border).
Before that time, properties are independent of $N$. This can be seen in the lower panel of Fig.~\ref{fig_p095_scaled_withN}, where
one observes that $p_\alpha(s,t,N)$ does not change keeping $t$ fixed as $N$ is varied, except for the region of very large domains where border finite-size effects start to be appreciable.
Such border finite-size effects are present also in the SLR case but, besides that, the slow decay of the interactions
introduces other finite-size effects which, at variance with the border ones, are present at any time. Basically, since a spin interacts with
any other, the number of the latter makes a difference. We will call these the {\it bulk finite-size effects}. 
An example can be seen in the inset of the upper panel of Fig.~\ref{fig_p095_scaled_withN}, where one observes that curves at fixed
$t$ do not collapse as $N$ is varied, and there is a systematic drift also for 
values of $s\ll N$. At a given time $t$, the border finite-size effects vanish in the thermodynamic limit, which is not true for the bulk ones.

Hence bulk finite-size effects must be taken into account in the scaling form. Indeed, it is shown in the main plot of the upper panel of 
Fig.~\ref{fig_p095_scaled_withN} that curves of $p_\alpha(s,t,N)$ at a fixed time and different $N$ can be collapsed, implying 
\begin{equation}
    p_\alpha(s,t,N)=N^{b_\alpha}\langle s_\alpha(t,N)\rangle ^{-1}\,h\left ( \frac{s}{N^{-b_\alpha}\langle s_\alpha(t,N)\rangle}\right ),\quad \mbox{for fixed }t,
    \label{scalpN}
\end{equation}
where $h$ is another scaling function. 
Notice that also in this case the very large domains with $s\lesssim N$ do not obey scaling
due to the border finite-size effects. It is immediate to check that, given the fat tail of $p_\alpha(s,t,N)$ discussed
above, the exponent $b_\alpha$ entering Eq.~(\ref{scalpN}) must be the same as the one regulating the large-$s$ decay of 
$p_\alpha(s,t,N)$ introduced before in Sec.~\ref{overall} and, in fact, the collapse observed in Fig.~\ref{fig_p095_scaled_withN} is obtained using the
same value $b=0.387$ already found previously.
The dependence of this exponent on $\alpha $ will be discussed soon.

Combining the effect of changing $t$ and $N$, expressed by the forms~(\ref{scalpt}) and~(\ref{scalpN}), one arrives at
\begin{equation}
    p_\alpha(s,t,N)=L^{-1}_\alpha(t,N)\,f\left (\frac{s}{L_\alpha(t,N)}\right ),
    \label{scalpfull}
\end{equation}
where
\begin{equation}
    L_\alpha(t,N)=\frac{\langle s_\alpha(t,N)\rangle}{N^{b_\alpha}}
    \label{eqL}
\end{equation}
has the meaning of a domains size measured in $N$-dependent units such that $L_\alpha(t,N)$ turns out to be size-independent (however we still keep the argument $N$ because, as we will see later, some residual $N$-dependence 
is present for finite $N$). 

\begin{figure*}
	\centering
	\rotatebox{0}{\resizebox{0.49\textwidth}{!}{\includegraphics{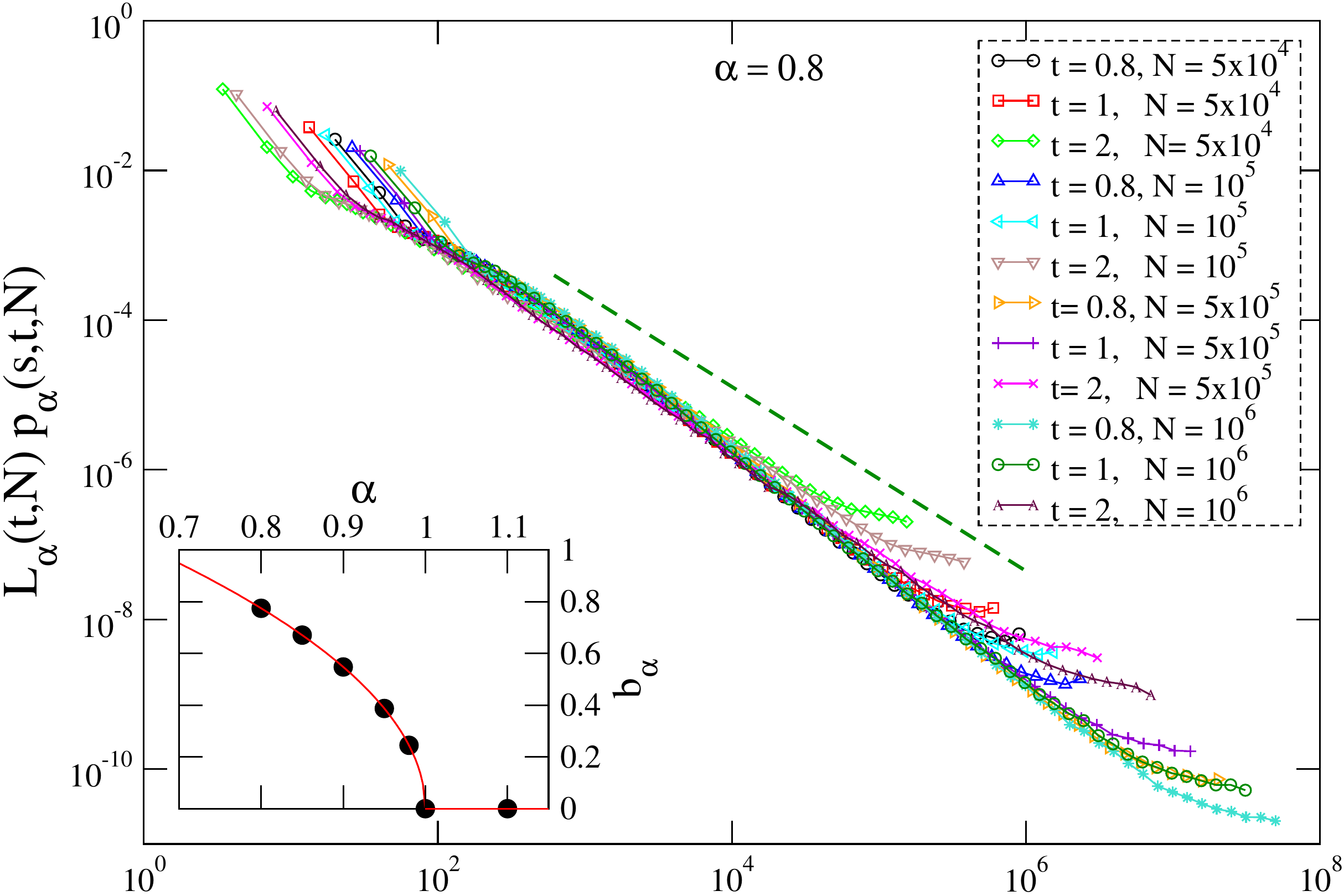}}} 
\vspace{0.3cm} \hspace{0.04cm}
 	\rotatebox{0}{\resizebox{0.49\textwidth}{!}{\includegraphics{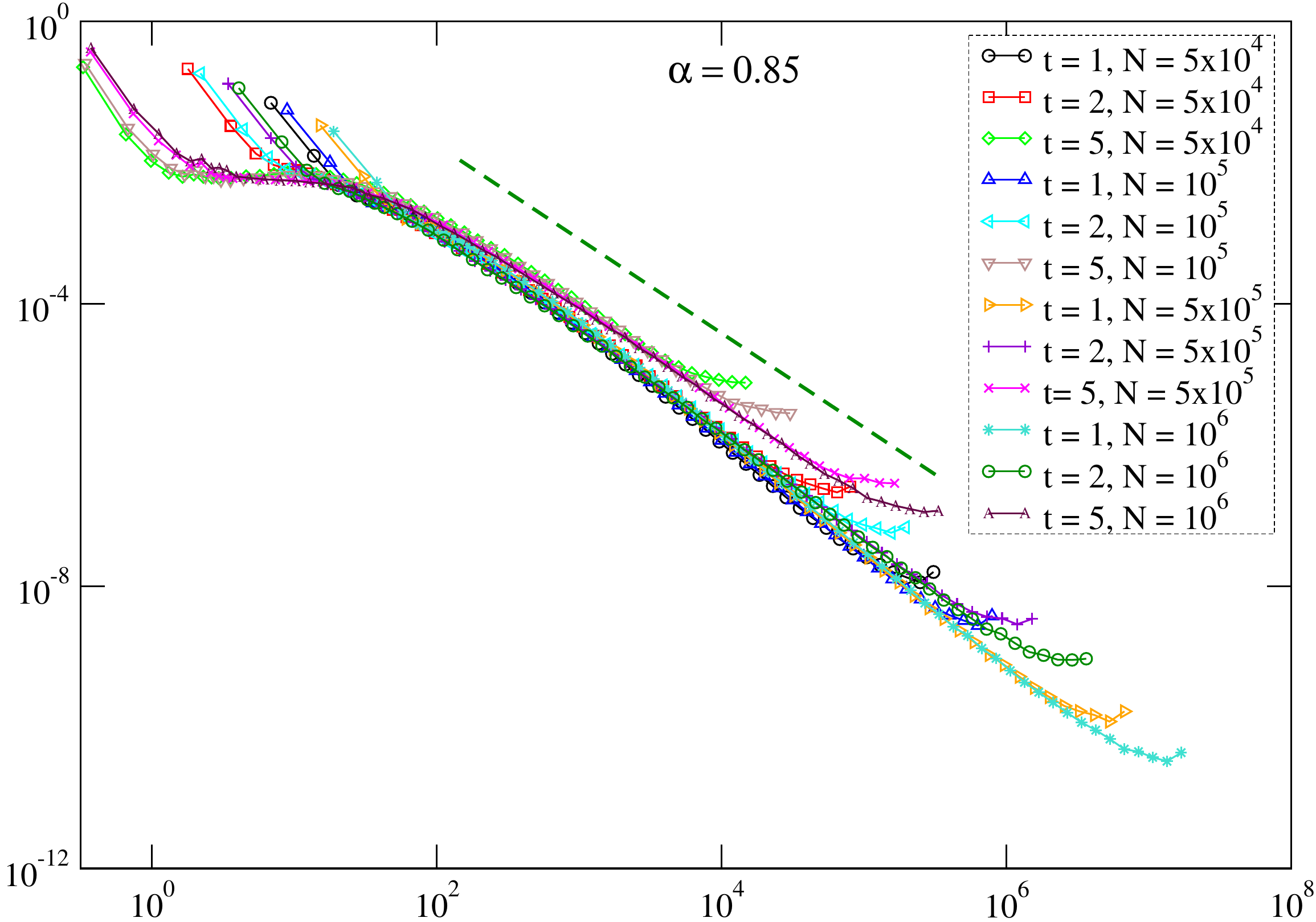}}} 
	\rotatebox{0}{\resizebox{0.49\textwidth}{!}{\includegraphics{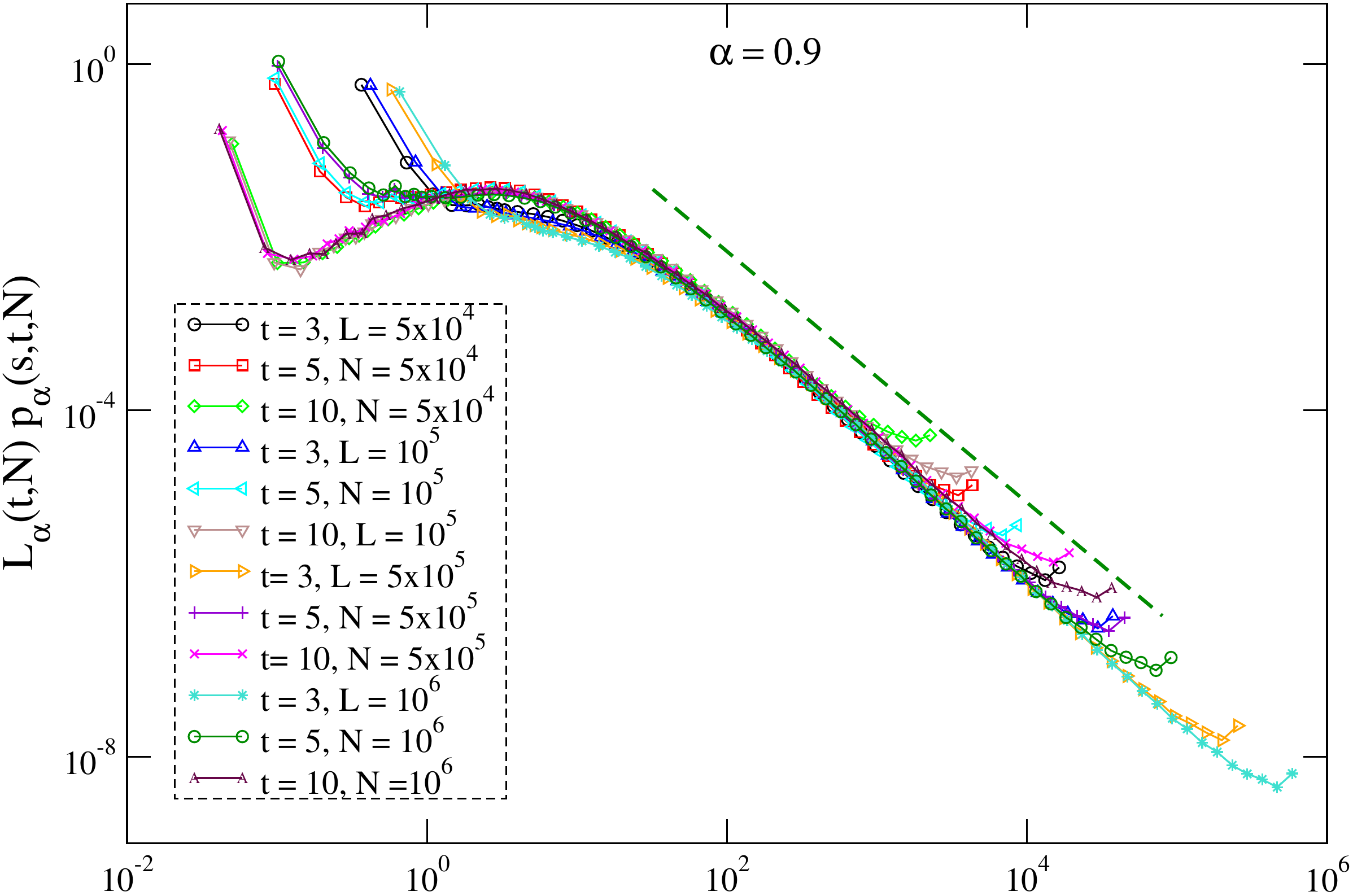}}}
 \vspace{0.3cm} \hspace{0.04cm}
	\rotatebox{0}{\resizebox{0.49\textwidth}{!}{\includegraphics{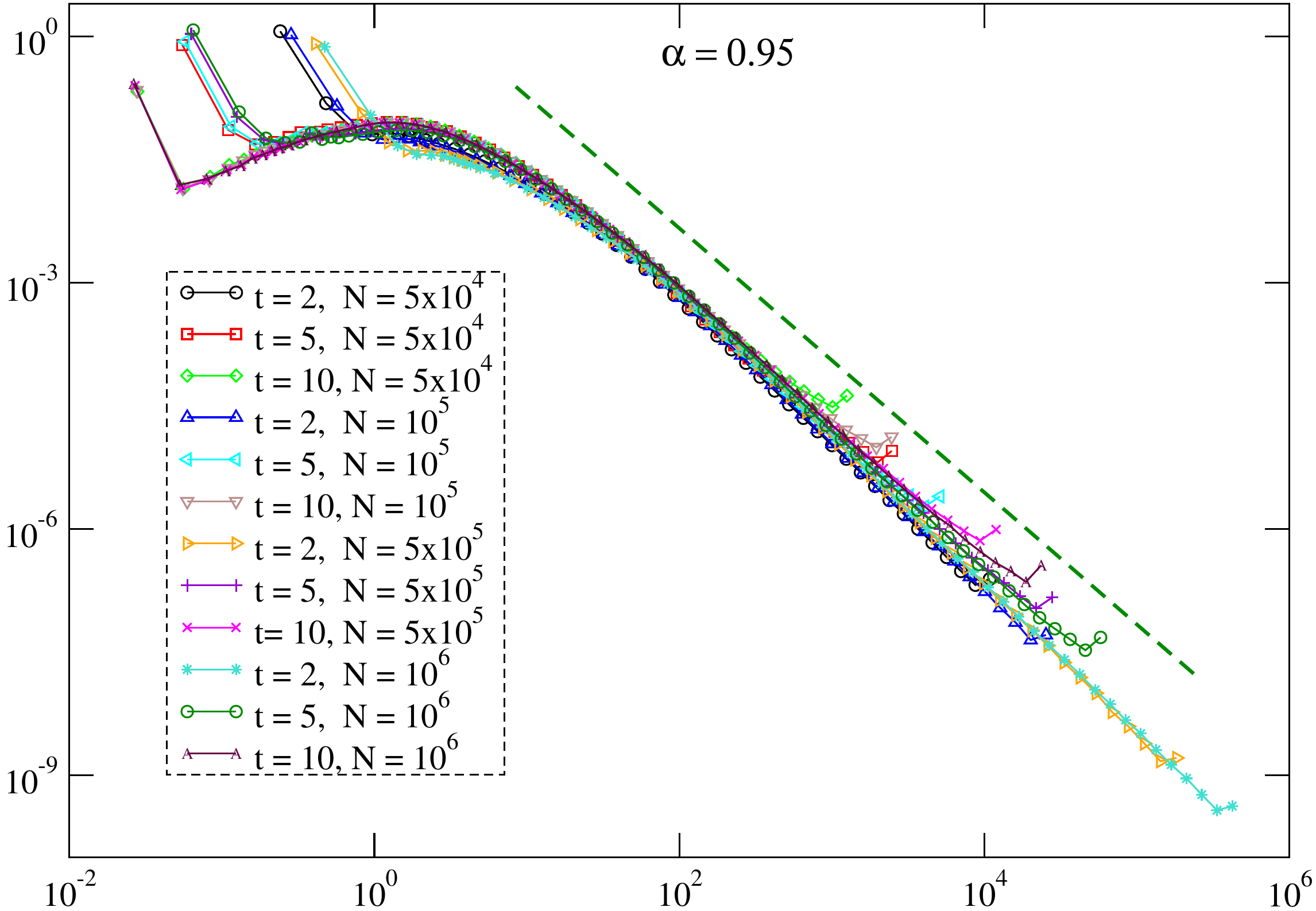}}} 
 	\rotatebox{0}{\resizebox{0.49\textwidth}{!}{\includegraphics{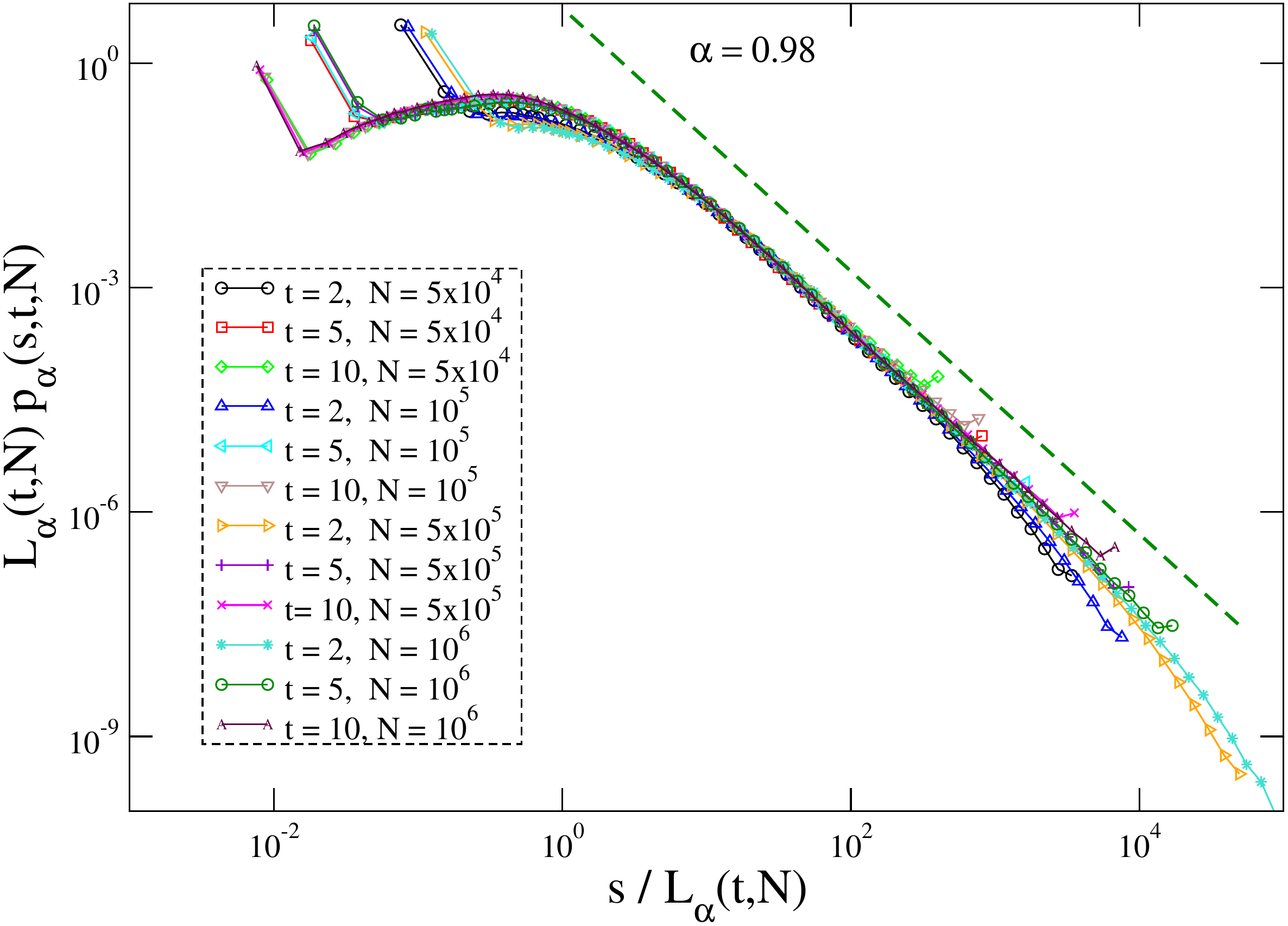}}}
    \hspace{0.04cm}
 	\rotatebox{0}{\resizebox{0.49\textwidth}{!}{\includegraphics{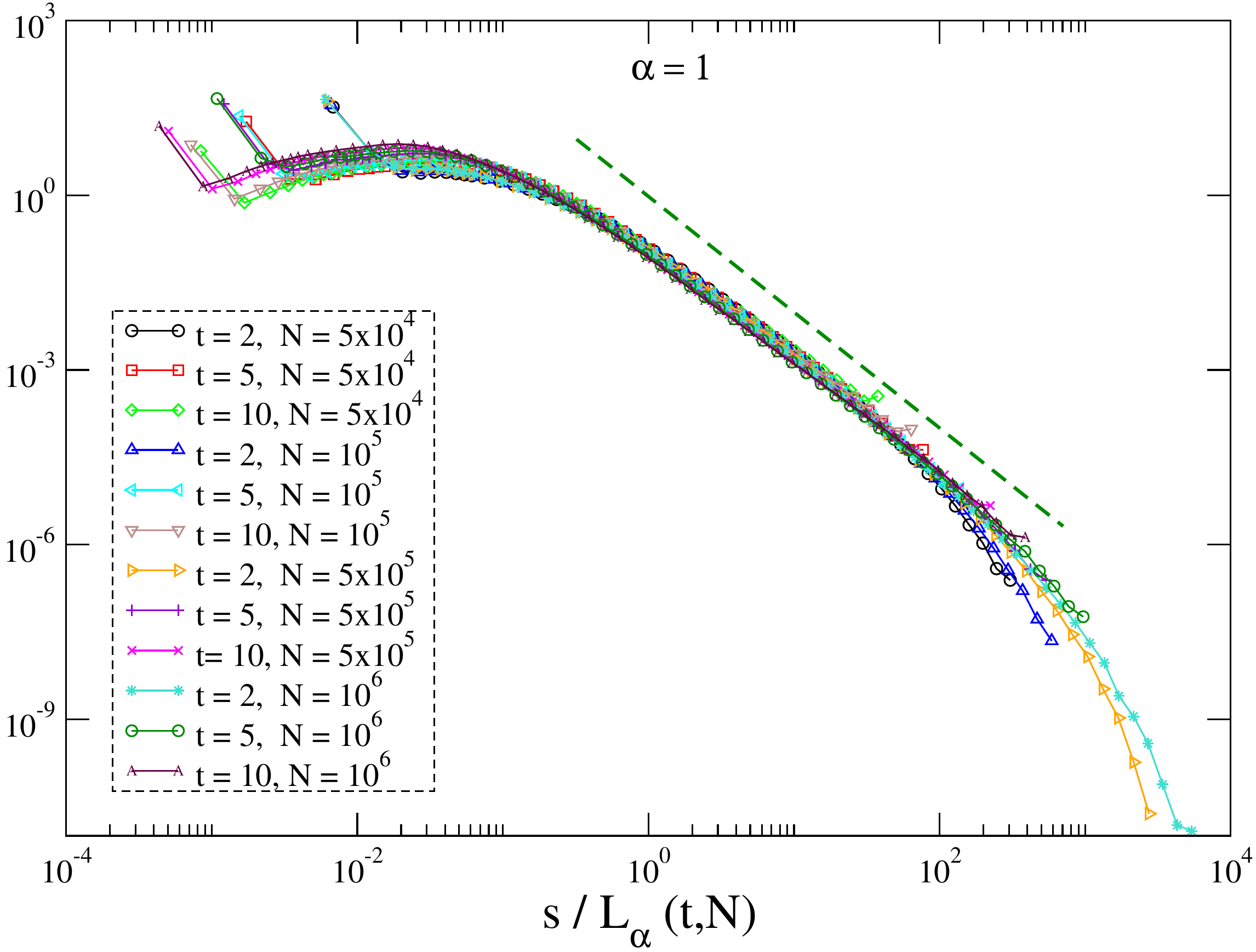}}}   
  \caption{$L_\alpha(t,N) \,p_\alpha(s,t,N)$ is plotted against $\frac{s}{L_\alpha(t,N)}$, for different values of $t$
 and $N$ (see key). Different panels refer to different values of $\alpha$, as indicated. The green dashed lines are the behaviours $\propto x^{-(2-b)}$, with $b_\alpha$ given in Eq.~(\ref{eqb}). In the inset of the upper-left panel the value of the $b_\alpha$ exponent obtained by best-fit procedure is shown (heavy circles) and compared with the law~(\ref{eqb}) (continuous red line).}
	\label{fig_scalpfull}
\end{figure*}

In Fig.~\ref{fig_scalpfull} we show the data collapse of the $p_\alpha$ curves as $t$ and $N$ are both changed, for 
several values of $\alpha$ (in different panels), fully confirming the scaling form~(\ref{scalpfull}). The exponent
$b$ has been obtained with the best fit procedure which, once performed for all the values
of $\alpha$ considered, leads to values very well consistent with the form
\begin{equation}
    b_\alpha=\left \{ \begin{array}{ll}
       \sqrt{3(1-\alpha)},  &  \quad \mbox{for}\,\, \alpha \le 1\\
        0, & \quad \mbox{for}\,\, \alpha >1, 
    \end{array}
    \right .
    \label{eqb}
\end{equation}
as it is shown in the inset of the upper-left panel of
Fig.~\ref{fig_scalpfull}.
For $\alpha=1$
a logarithmic correction is expected which, however, with the present quality of the data, is not visible.
Clearly, values of $b_\alpha$ slightly different from the ones given by the putative form~(\ref{eqb}) could also provide a good data collapse; however notice that in the investigated range of $\alpha$, $0.8\le \alpha \le 1$,
the exponent $b_\alpha$ changes significantly, going from $b_{\alpha=1}=0$ to $b_{\alpha=0.8}\simeq 0.77$.
Similar disclaimers apply to the estimation of other parameters (such as 
$t^*(\alpha)$, see Sec.~\ref{growthlaw}), because the important and ever-present border finite-size effects make the values of $N$ attainable in our simulations quite insufficient to control the scaling in the truly very large-$N$ limit.
This issue will 
be further discussed later on.
Despite this, in the range of $N$ and $t$ that can be accessed in the simulation, forms such as the one in Eq.~(\ref{eqb}) describe quite well the dataset and, as we will see, can be used to infer the true large-$N$ behaviour.

\subsection{Growth law} \label{growthlaw}

\begin{figure}[h]
	\centering
	\rotatebox{0}{\resizebox{0.45\textwidth}{!}{\includegraphics{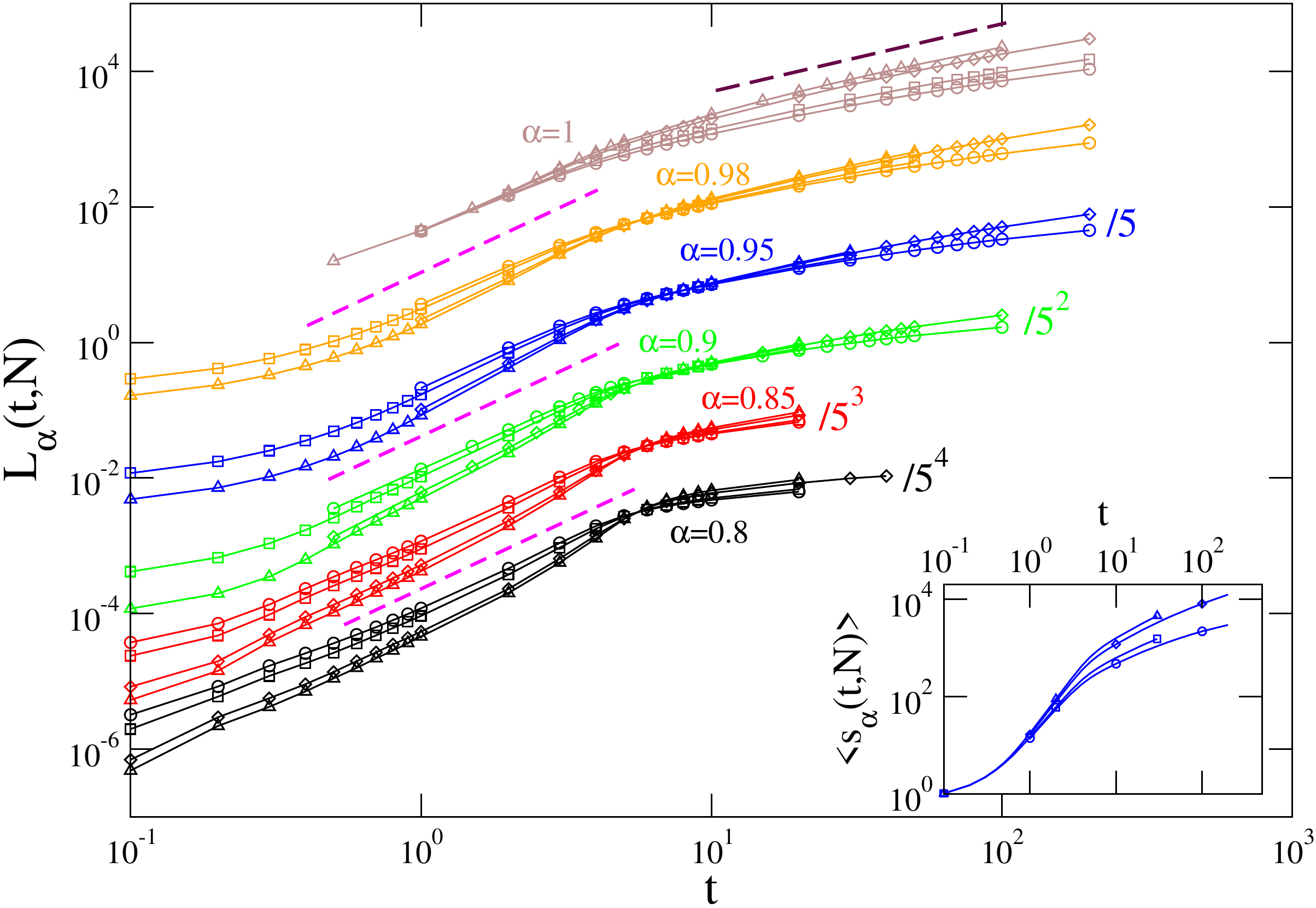}}} 
  \caption{$L_\alpha(t,N)$ is plotted against time, for various system sizes and
  different values of $\alpha$ (from bottom to top: black $\alpha=0.8$, red $\alpha=0.85$, green $\alpha=0.9$, blue $\alpha=0.95$, orange $\alpha=0.98$, brown $\alpha=1$, as also indicated in figure). Curves for $\alpha=0.95, 0.9, 0.85$ and $0.8$ are shifted downwards by factors
  $5, 5^2, 5^3$ and $5^4$, respectively, as also indicated in the figure. The symbols circles, squares, diamonds and triangles refer to the   system sizes $N=5\cdot 10^4, 10^5, 5\cdot 10^5$ and $10^6$, respectively.   The dashed magenta and maroon lines are the behaviours $\propto t^2$ and $\propto t$, respectively. In the inset the same data of the main figure for $\alpha=0.95$ are plotted without the $N^b$-rescaling namely, looking at Eq.~(\ref{eqL}),
  plotting $\langle s_\alpha (t,N) \rangle$.}
	\label{fig_L}
\end{figure}

In this section we discuss the growth law
of $\langle s_\alpha(t,N)\rangle$ (or, equivalently, of $L_\alpha (t,N)$, see Eq.~(\ref{eqL})), which was briefly
considered already in Sec.~\ref{overall}.
This quantity is shown in Fig.~\ref{fig_L} where, in order to separate the cases with different values of $\alpha$, we have
shifted some curves vertically, as explained in the caption. 

From this figure some conclusions can be drawn. Firstly, the two different regimes previously indicated as I and II, are clearly visible
for any choice of $\alpha$, the crossover among them occurring around $t^*_\alpha\simeq 6-7$ for any value of $\alpha$. 
Secondly, it is true that going from $\langle s_\alpha(t,N)\rangle$ to $L_\alpha(t,N)$ removes, to a good extent, the dependence
on $N$ for $t\gtrsim t^*_\alpha$. Indeed, for a fixed $\alpha$, all curves for different $N$ cross around the crossover time from regime I to II
while there is a systematic increase of the curves for $\langle s_\alpha(t,N)\rangle$ with $N$, at any time, as it can be appreciated in the
inset for $\alpha=0.95$. Notice that having $L$ (approximately) independent of $N$ implies, through Eq.~(\ref{eqL}),
that $\langle s_\alpha(t,N)\rangle$ diverges in the thermodynamic limit 
also at \textit{finite} times.
Despite that, however, some $N$ dependence is left, as can be seen because the curves really collapse only when they cross. Let us anticipate that we ascribe this to border finite-size effects.
The data for the largest value of $N$ ($N=10^6$) provides a good indication that the growth-law is
\begin{equation}
    L_\alpha(t,N)\propto t^2 \quad \quad \mbox{(regime I)}
\end{equation}
in regime I. This is better observed for smaller values of $\alpha$ because $t^{\rm micro}_\alpha$ is smaller and the regime sets in already at very short times.

It is difficult, on the other hand, to establish the growth-law in regime II. Indeed, by looking again at the largest system size, one has 
a good indication of a ballistic growth $L_\alpha (t,N)\simeq t$ for the largest values of $\alpha$, but the slope of the curves at late times
sensibly decreases as $\alpha $ is decreased. For example, fits of the slope around $t=10$ give $1.14, 1.05, 0.99, 0.91, 0.76, 0.53$ for $\alpha=1, 0.98, 0.95, 0.9, 0.85, 0.8$, respectively. These values are probably not good estimators of the growth-law exponent in regime II, because it is difficult to locate correctly the crossover
between regimes I and II, and because the slopes tend to decrease slightly as time grows larger and larger. Despite that, the number above clearly speak of slower growth at small $\alpha$. 
It is true that, for fixed $\alpha$, the slope increases with $N$, but for the cases with small $\alpha$ it seems unlikely, by looking data in this way,  that this effect might
be able to reinstate the ballistic growth observed for the largest values of $\alpha$. Clearly, simulations with much larger sizes than the one considered here could help to resolve this question, 
but it must be recalled that the computational time increases as $N^2$ and much larger values of $N$ are not feasible. 

This is the information we can honestly get from the present data. What follows, instead, regards their physical interpretation, which can be two-fold. On the one hand, adhering to a strict data-oriented interpretation, one could conclude, looking at Fig.~\ref{fig_L} that data for fixed $\alpha$
and different sizes $N$ do not actually collapse and that, changing $\alpha$ at fixed $N$, the growth law is different for different $\alpha$. On the other hand, such interpretation implies some facts that clash both with physical intuition and, in a broader sense, with what is known in coarsening phenomena at a general level. Indeed, data in Fig.~\ref{fig_L}, taken as they are, would lead to the conclusion that growth is slower at smaller $\alpha$, a fact conflicting with the physical expectation that 
a larger cooperation (i.e. smaller $\alpha$) can only speed up the process or, at most, leave it unchanged. This is indeed what happens for WLR, both for quenches to finite temperatures or to $T=0$. 

Furthermore, the lack of collapse of curves with equal $\alpha$ in Fig.~\ref{fig_L} contradicts scaling which however is a common feature  in coarsening phenomena. Indeed, Fig.~\ref{fig_scalpfull} shows that 
scaling is quite well obeyed by the $p_\alpha$ and this suggests that, due to that, one should expect a scaling form for $L_\alpha(t,N)$ of the type 
\begin{equation}
    L_\alpha(t,N)=\ell \left(\frac{t}{t^*_\alpha} \right ), \quad \quad \mbox{for } N\to \infty
    \label{scalL}
\end{equation}
where $t^*_\alpha$ is some characteristic time setting the timescale of the ordering process, and $\ell$ is a scaling function. Recalling what discussed in  Sec.~\ref{overall}, the only characteristic times are the microscopic time $t^{\rm micro}_\alpha$ and the crossover time $t^*_\alpha$; since scaling is expected in the large-time domain it is natural to identify the characteristic time appearing in Eq.~(\ref{scalL}) with the crossover time $t^*_\alpha$ previously introduced in Sec.~\ref{overall}, hence the use of the same symbol also here. The form~(\ref{scalL}) should be obeyed at least in the asymptotic time domain, i.e. in regime II (but before border finite-size effects set in). Eq.~(\ref{scalL}) embodies the already discussed fact that, for large $N$, $L_\alpha(t,N)$
should be independent on $N$, $\lim _{N\to \infty}L_\alpha (t,N)=L_\alpha(t,\infty)$ and that the $\alpha$ dependence should enter $L_\alpha(t,\infty)$ through a typical time $t^*(\alpha)$, forming
a ratio as in Eq.~(\ref{scalL}). 

This leads to a different, more physically-oriented interpretation, to which we are more committed. According to this interpretation, in the large-$N$ limit $L_\alpha (t,N)$ is actually independent of $N$
and the asymptotic growth law (i.e. in regime II but before border finite-size effects take place) is ballistic, i.e. 
\begin{equation}
    L_\alpha(t,N)=L_\alpha(t)\propto t
\end{equation}
for any value of $\alpha$. This particular growth-law is suggested by an analytical argument that will be discussed in Sec.~\ref{analytical}. According to this interpretation, all the deviations from this expected (for sufficiently large $N$) behaviour which are observed in Fig.~\ref{fig_L}, are due to the border finite-size effect and are bound to disappear if we were able to perform simulations of unfeasible systems with much larger values of $N$.

\begin{figure}[h]
	\centering
	\rotatebox{0}{\resizebox{0.45\textwidth}{!}{\includegraphics{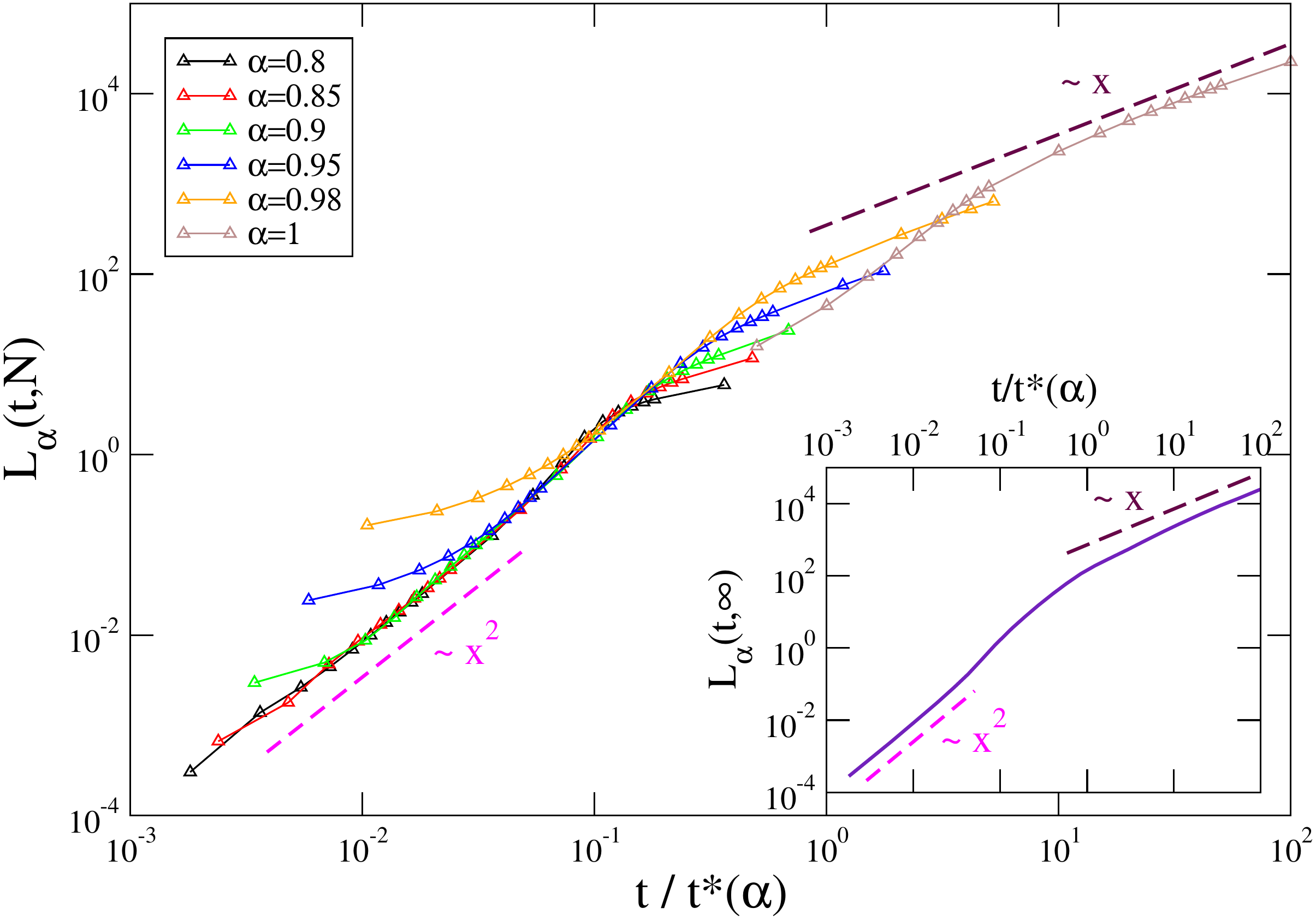}}} 
  \caption{$L_\alpha(t,N)$ is plotted against $\frac{t}{t^*(\alpha)}$, for $N=10^6$, and
  different values of $\alpha$ (see key).  The dashed magenta and maroon lines are the behaviours $\propto x^2$ and $\propto x$, respectively. In the inset the mastercurve expected for 
  $N\to \infty$ is shown. This has been obtained from the data in the main figure, by keeping only the points where curves   for different $\alpha$ collapse (with some interpolation when needed).}
	\label{fig_scale_L}
\end{figure}

In Fig.~\ref{fig_scale_L} we have tested our hypothesis by using $t^*_\alpha$ as an adjustable parameter leading to data collapse (for clarity we draw only the data for the largest system size). Although it is impossible to superimpose curves for different $\alpha$, a nice collapse can be obtained in regime I with the choice
\begin{equation}
    t^*_\alpha=e^{6(1-\alpha)^{1/4}}.
\end{equation}
This form, like Eq.~(\ref{eqb}), with a non-analytic behaviour at $\alpha=1$, clearly indicates that the nature of the ordering process changes abruptly for $\alpha >1$. 

If the present scaling approach is correct, the limiting curve on which the data collapse (which has been reconstructed in the inset of Fig.~\ref{fig_scale_L}, for clarity) represents the thermodynamic limit, whereas the portion of data outside this mastercurve has to be interpreted as being affected by strong finite-size effects. The observed behaviour of $L_\alpha(t,N)$ at different $N$ is consistent with this, since inspection of Fig.~\ref{fig_L} shows that, for each $\alpha$, the small-time part of the curves tends to decrease (towards the behaviour $\propto t^2$), while the large time part tends to increase (towards the behaviour $\propto t$). The mastercurve describes a crossover phenomenon between an early and a late regime, with the limiting behaviours.

\begin{equation}
    \ell(\tau)\propto \left \{ \begin{array}{ll}
\tau^2,         &\mbox{for } \tau\ll 1   \\
 \tau        &\mbox{for } \tau \gg1. 
    \end{array}
    \right .
\end{equation}
This form, in turn, would imply that the growth in regime II is ballistic for all values of $\alpha$, in the thermodynamic limit, in agreement with the analytical calculation that we discuss below.

\section{Analytical approach} \label{analytical}

As we have discussed in Sec.~\ref{model}, at $T=0$ only the sign of the local
fields $h_i$ determines the evolution, meaning that we can reduce such fields to boolean variables $h_i=\pm 1$. Then one can 
cast the transition rates~(\ref{GlauberRates}) as
\be
w(s_i)=\frac{1}{2}\left [1-s_i\tanh(\beta h_i)\right ]=
\frac{1}{2}\left [1-s_i\,\mbox{sign} (h_i)\right ]\simeq 
\frac{1}{2}\left (1-s_i h_i\right ),
\label{approxRates}
\end{equation}
where in the last passage we have approximated the sign of the local field with the field itself.
This approximation is valid in the interiors of domains when their size is sufficiently large (see Eqs.~(\ref{locfield}-\ref{eqKac}), i.e. for $t>t_\alpha^{\rm micro}$. It is 
not correct on interfaces, where $h_i$ changes sign, but the number of such points
is negligible in the limit of large domains.


Starting from a master-equation, it is straightforward~\cite{Glauber1963} to obtain the evolution equation for correlators of the general form
\begin{equation}
    \frac{d}{dt}\langle s_{i_1}(t)\cdots s_{i_n}(t)\rangle=\left \langle 
    s_{i_1}\cdots s_{i_n}\sum _{m=1}^nw(s_{i_m}(t)) \right \rangle.
    \label{general}
\end{equation}
For $n=1$, using the transition rates~(\ref{approxRates})
one has
\be
\frac{d}{dt}\langle S_i(t)\rangle=-\langle S_i(t)\rangle+ \sum_{r=1}^{N/2}J(r)\sum_{k=i\pm r}\langle S_k(t)\rangle =0
\label{eqmag}
\ee
where, in the last passage, we have assumed space translation invariance, which
is obeyed for $N\to \infty$.
This shows that, in this limit, the magnetisation cannot develop.

Similarly, letting $n=2$ in Eq.~(\ref{general}) one easily arrives at

\begin{eqnarray}
\frac{d}{dt}\langle s_i(t)s_j(t)\rangle&=&-2\langle s_i(t)s_j(t)\rangle \nonumber \\
& & +
\sum _{r=1}^{N/2}J(r)\left [\sum_{k=i\pm r}\langle s_j(t)s_k(t)\rangle
+\sum _{q=j\pm r}\langle s_i(t)s_q(t)\rangle \right ] \nonumber \\
&=& -2\langle s_i(t)s_j(t)\rangle \nonumber \\
& & +2\sum _{r=1}^{N/2}J(r)\left [ \langle s_i(t)s_{j-r}(t)\rangle+\langle s_i(t)s_{j+r}(t)\rangle \right ],
\label{eqc1}
\end{eqnarray}
having assumed space translation invariance again. Introducing $C(r,t)=\langle s_i(t)s_{i+r}\rangle$, Eq.~(\ref{eqc1}) reads
\be
\frac{d}{dt}C(r,t)=-2C(r,t)+2 \sum_{\ell=1}^{N/2}J(\ell)\left [C([[ r-\ell]],t)+C([[r+\ell]],t)\right ],
\label{eqc2}
\ee
where the symbol $[[n]]$ is used to take into account periodic boundary conditions, i.e.
\be
[[n]]=\left \{ \begin{array}{lll}
|n|, & \mbox{if}& |n|\le N/2 \\
N-|n|, & \mbox{if} &|n|>N/2. 
\end{array} \right .
\ee 

\begin{figure}[h]
	\centering
	\rotatebox{0}{\resizebox{0.45\textwidth}{!}{\includegraphics{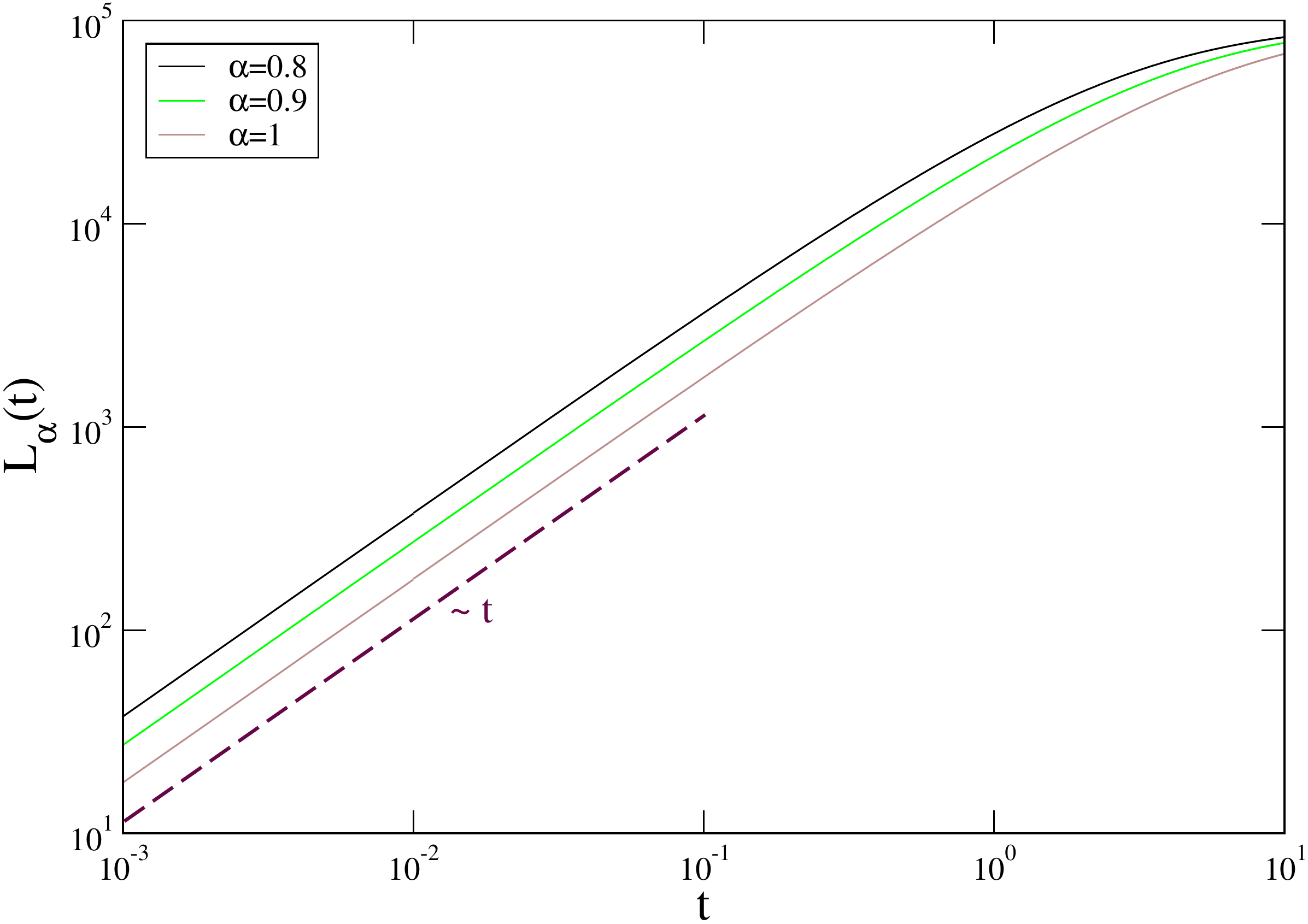}}} 
  \caption{$L_\alpha(t)$ computed as in Eq.~(\ref{eqR}) after solving numerically Eq.~(\ref{eqc2}) is shown for $N=10^5$ and different values of $\alpha$ (see key). The dashed line is the ballistic behaviour $L_\alpha(t)\propto t$.}
	\label{fig_ising_LR_eqC}
\end{figure}

Eq.~(\ref{eqc2}) can be easily solved numerically, starting from a fully disordered initial condition
with $C(r,0)=\delta _{r,0}$.
From the knowledge of $C(r,t)$, one can extract a characteristic size $L_\alpha (t)$ as
\be
L_\alpha(t)=4\,\frac{\sum _{r=0}^{N/2}r\,C(r,t)}{\sum _{r=0}^{N/2}C(r,t)},
\label{eqR}
\ee
where factor 4 has been introduced in order to have $L_\alpha=1$ on a fully ordered configuration.
This quantity is plotted in Fig.~\ref{fig_ising_LR_eqC}. Data show a ballistic growth of $L_\alpha(t)\propto t$, before saturation to a final value $L_\alpha (t\gg 1)\simeq N$ takes place. Deviation from ballistic behaviour takes place earlier upon decreasing $\alpha$. For the values of $\alpha$ considered in Fig.~\ref{fig_ising_LR_eqC} this occurs around $L_\alpha(t)\simeq N/50$, meaning that such effects sets in quite early even for relatively large systems, and are, therefore, particularly severe.  
One can use this estimation to assess if the data from our numerical simulations of the original model are free from border finite-size effects. In doing that one discovers that for the largest system size considered ($N=10^6$), only for $\alpha =1$ and, marginally, for $\alpha =0.98$ there exists a border finite-size free time window where the ballistic growth can be observed. According to this interpretation, the slower growth we observe for other values of $\alpha$ is due 
to the flattening of the growth-law caused by the finite size, similarly to what found in this analytical approach.

In summary, the present analytical argument establishes a ballistic growth at late times.  
A first important difference with the numerical simulations is that here magnetisation does not develop, see Eq.~(\ref{eqmag}). Instead, in the original model, a small initial up-down unbalance grows, as it happens in mean-field.
It must also be recalled that in the numerical simulations we restrict the statistics to the configurations containing domains. Configurations with a relatively large magnetisation will never develop domains or, if they do so, such domains will be eliminated quite early and the corresponding configurations soon exit the statistics. This being said, we expect the analytical solution
to describe correctly the behaviour of the original model at late times, namely in regime II, before
border finite-size effects become relevant. This is true not only because, as already discussed, the approximation involved in the transition rates~(\ref{approxRates}) is reliable only in this
limit, but also because at late times, configurations which still have domains have necessarily a 
tiny magnetisation (this has been checked numerically) and, as we argued, it is precisely the
development of magnetisation a main source of difference between the analytical approach and the simulations.  

\section{Conclusions} \label{concl}

In this paper we have considered the phase-ordering kinetics of the one-dimensional Ising model with long-range interactions decaying with distance as $J(r)\propto r^{-\alpha}$, with $\alpha \le 1$. In a previous paper~\cite{iannone2021} it was shown that, in this process, the statistical non-equilibrium ensemble splits into two different kinds of trajectories, those which develop coarsening domains, and those which do not so and behave similarly to mean field. Restricting the observation on the former ones, in the present study we have performed numerical simulations to compute the probability distribution $p_\alpha (s,t,N)$ of the domain sizes, from which the growth-law of the average domains size $\langle s_\alpha(t,N)\rangle $ can be inferred.

Our analysis shows that $p_\alpha(s,t,N)$ takes a scaling form, Eq.~(\ref{scalpfull}), where the $t$ and $N$ dependence can be accounted for in terms of a growing length $L_\alpha(t,N)$. This suggests that a dynamical scaling symmetry similar to the one observed in coarsening systems with short-range interactions is at work where, in addition, the long-range nature introduce non-trivial dependencies on the number $N$ of spins. Such finite-size effects can be grouped into two classes: the former refers to the size-dependence of observable quantities affecting the process at any time. We termed these 
bulk finite-size effects. Next to these, there are border finite-size effects, which are felt only at late times when the characteristic size of the domains becomes comparable with the system size and are akin to the ones observed in systems with short-range interactions.

Numerical data show a superballistic growth $L_\alpha (t,N)\propto t^2$ at short times, and a different behaviour at long times. Complementing the numerical observations with an analytical treatment of the model which is expected to be accurate in the late time-domain, we argue that the asymptotic growth-law (before the border effect becomes important) is ballistic, similarly to what happens for $\alpha >1$. This behaviour is expected to be observed in the thermodynamic limit $N\to \infty$, or, in a finite system, when the characteristic size $\langle s_\alpha (t,N)\rangle$ of the growing domain is much smaller than $N$. However, the analytical calculations of Sec.~\ref{analytical} shows that this condition is particularly severe, requiring $\langle s_\alpha (t,N)\rangle \lesssim N/50$, asking for numerically unfeasible system sizes unless $\alpha $ is sufficiently close to $\alpha=1$. 

The results presented in this paper represent the first contribution to the quantitative understanding of phase-ordering in a ferromagnetic model with SLR interactions. Several open questions remain to be addressed. First of all, still considering the one-dimensional model, the role of temperature fluctuations, in quenches to $T>0$ is fully unexplored. In the WLR case with $\alpha >1$, in fact, it is
known~\cite{CLP_review,CLP_epl,CLP_lambda} that the ballistic regime leaves room for a different, $\alpha$-dependent growth law. Similarly, the phase-ordering occurring in dimensions larger than one has never been investigated. These issues remain open to further research studies.

\section*{Acknowledgement}

All numerical simulations presented here were done on the Zeus as well as EPYC HPCs of Coventry University in the United Kingdom.

\bibliographystyle{elsarticle-num} 

\bibliography{strongLR}

\end{document}